\documentstyle[12pt,aaspp4]{article}

\lefthead{Bekki Kenji}
\righthead{Starburst in multiple galaxy mergers}

\begin{document}
\title{Starburst in multiple galaxy mergers}

\author{Kenji Bekki} 
\affil{Division of Theoretical Astrophysics,
National Astronomical Observatory, Mitaka, Tokyo, 181-8588, Japan}

\begin{abstract}

We numerically investigate stellar and gaseous dynamical evolution
of mergers between five identical late-type disk galaxies
with the special emphasis on star formation history and chemical
evolution of multiple galaxy mergers.
We found that multiple encounter and merging can trigger   
repetitive massive  starbursts  (typically $\sim$ 100 $M_{\odot}$ 
${\rm yr}^{-1}$)
owing to the strong  tidal disturbance
and the resultant gaseous dissipation during merging.
The magnitude of the starburst is found to depend on initial
virial ratio (i.e., the ratio of total kinematical
energy to total potential energy) 
such that the maximum star formation rate
is larger for the  merger with smaller virial ratio. 
Furthermore we found that
the time interval between the epochs of the triggered starbursts
is longer for the merger with the larger virial ratio. 
The remnant of a multiple galaxy merger with massive starbursts
is found to have  
metal-poor gaseous halo that is formed by tidal stripping
during the merging.
We accordingly suggest that metal-poor gaseous halo 
in  a field elliptical galaxy is a fossil record of the past
multiple merging events for the galaxy.

\end{abstract}

\keywords{galaxies: starburst -- galaxies: ISM -- 
galaxies: elliptical and lenticular, cD -- galaxies: formation --
galaxies:
interaction -- galaxies: structure 
}

\section{Introduction}

Major galaxy merging between two gas-rich
spirals is observationally suggested  to play
vital roles in the formation and the evolution
of galaxies; Formation of elliptical galaxies 
(e.g., Schweizer 1986; Kormendy \& Sanders 1992; Bender 1996),
QSO formation (e.g., Stockton 1997),
formation
of fine structure of elliptical galaxies (e.g., Schweizer 1997),
triggering massive starburst and the resultant  
formation of ultra-luminous infrared galaxies
(ULIRGs) (e.g., Sanders et al. 1988; Sanders \& Mirabel 1996),
and globular cluster formation (e.g., Whitemore et al. 1995).
Accordingly a growing number of theoretical
studies have tried to answer the questions
related to physical processes of major galaxy merging. 
These include, for example, 
the formation of elliptical galaxies (Toomre 1977;
Barnes 1988, 1992; 
Barnes \& Hernquist 1992a, 1996; 
Hernquist 1992, 1993),
physical mechanism for triggering massive starburst 
(e.g., Mihos \& Hernquist 1996; Gerritsen \& Icke 1997),
and formation and evolution of 
ULIRGs (e.g., Bekki , Shioya, \& Tanaka 1999).
In comparison with the above theoretical studies on $pair$ galaxy merging,
there are a relatively small number of theoretical studies
on the  multiple galaxy merging between more than two galaxies,
though multiple merging can be important for some aspects
of galaxy evolution, such as the evolution of a small group
of galaxies (Barnes 1989; Hickson 1997).
This is primarily because there are only a few observational
results showing  unambiguous and direct 
evidences  that  multiple galaxy merging is not rare 
in the Universe.

Theoretical studies on multiple galaxy merging can be divided
into the following two categories: (1) Formation of elliptical
galaxies and (2) Evolution of compact group of galaxies.
Concerning the first issue, Barnes (1989) first demonstrated that 
multiple interaction and merging of a compact group (such as
Hickson compact groups, hereafter referred to as HCGs)
results in the formation of a bright elliptical galaxy. 
Weil \& Hernquist (1994) revealed that multiple merging between
several stellar disks can successfully reproduce 
the small angle between rotation and minor axis
observed in elliptical galaxies.  Weil \& Hernquist (1996) 
furthermore discussed the difference in structural and kinematical
properties of merger remnants between pair mergers and multiple ones.
Concerning the second issue,
most of  studies have focused on the problems
on the time scale within which a compact group of galaxy
becomes a singe object  after successive merging (Carnevale et al. 1981;
Ishizawa et al. 1983; Barnes 1985; Ishizawa 1986; 
Mamon 1987, 1990; Navarro et al. 1987; 
Governato, Bhatia \& Chincarini 1991;
Bode et al. 1993;
Diaferio, Geller, \& Ramella 1994;
Athanassoula, Makino, \& Bosma 1997;
Hickson 1997
for a recent review).
Barnes (1985) demonstrated that a common dark halo of a compact
group can slow down the merging between the member galaxies
and consequently the lifetime of the group can be 5 $\sim$ 10
times longer than its apparent crossing time.
The roles of a  massive background dark halo in delaying
the merging time scale of a compact group 
have been confirmed by several numerical
studies (e.g., Navarro et al. 1987).  
Athanassoula et al. (1997) found that 
the time scale for which 
a group of galaxies can merge to form a single massive galaxy
can be larger than a Hubble time for some appropriate physical
conditions of the group.
They  therefore suggested
that the previously suggested puzzle on the existence of HCGs
(i.e., why HCGs can be observable nevertheless the crossing times
of them are very short  (typically $\sim$ 0.2 Gyr) and dynamical friction
should be so efficient that the galaxies in the groups can all merge  
in a time scale much smaller than a Hubble time ?) 
is not necessarily a problem.
Although these previous studies have paid much attention
to the structure and kinematics of the remnants of multiple mergers
and the merging history,  gaseous  dynamical process and star formation
history, both of which are important for understanding
the origin of
stellar populations of field elliptical galaxies and thus for
photometric and spectroscopic evolution of them, 
have not been yet investigated.

The purpose of this paper is to investigate star formation history
and chemical evolution of multiple galaxy mergers between
five identical gas-rich spirals.
We particularly investigate when and how massive starbursts,
which are demonstrated to occur in (pair) major mergers 
between two gas-rich spirals (Mihos \& Hernquist
1996), are triggered in multiple galaxy merging.
We furthermore investigate  two-dimensional (projected) distribution 
of metallicity both for stellar components and for gaseous ones
in merger remnants in order to show chemical properties
characteristics of the remnants of multiple galaxy merging. 
Based on  the central near infrared colors
of  elliptical galaxies with the sign
of recent merging,
Silva \& Bothun (1998) have demonstrated that
if elliptical galaxies are formed by merging,
the formation redshift is rather high.
Guided by this observational result,
we consider multiple merging between rather gas-rich disks with
the gas mass fraction of 0.2 (at higher redshift).
Star formation history and chemical evolution
of multiple mergers have not been investigated at all in previous
numerical and theoretical studies.
Thus the present results can provide some  implications
on the formation and the evolution of galaxies,
in particular, on the origin of field elliptical galaxies,
the evolution of compact group of galaxies, and the formation
of infrared luminous galaxies such as ULIRGs.
In the present paper, we furthermore consider 
that it is  important to investigate observationally what fraction
of ULIRGs are formed by multiple galaxy merging.
This is
because the fraction enables us to estimate (though indirectly)
the fraction of compact groups that are  now evolving
from ULIRGs into field elliptical galaxies.
We accordingly discuss the fraction of ULIRG and field elliptical
galaxies formed by multiple merging in the discussion section.

The layout of this paper is as follows.
In \S    2, we summarize  numerical models used in the
present study and describe in detail the methods for 
solving the star formation history and the chemical evolution
of multiple mergers.
In \S 3, we present numerical results on the time evolution
of morphology,  star formation history, and chemical  properties in
multiple  mergers. In \S 4, we discuss the strength of starbursts
triggered in multiple merging and the chemical properties
of the merger remnants. 
The conclusions of the preset study
are  given in \S 5.

\section{Model}

Our previous studies (Bekki \& Shioya 1998) have
already described in detail 
the initial conditions of merger progenitor disks,
the prescriptions of dissipative process,  the model for star formation,
that for chemical evolution of galaxy mergers,
and numerical method for solving dynamical evolution of stellar and gaseous
components.
Accordingly we describe only briefly the merger model in the present
study.
 We consider a multiple merger between five identical gas-rich spiral
galaxies without bulges. The disk mass and size 
are  assumed to be
exactly the same between disk galaxies in a multiple merger
and similar to  those of  the Galaxy.
Each of the disks has individual dark matter halo, and we do not
consider that a common dark halo can envelop a multiple merger
composed of five disks in the present study.
Although a common dark halo
can affect greatly the merging time scale in multiple galaxy merging
(Barnes 1985; Athanassoula et al. 1997),
we here do not consider this important effect of the common halo.

\subsection{Disk model}

 We construct  models of galaxy mergers between gas-rich 
 disk galaxies with equal mass by using Fall-Efstathiou model (1980).
 The total mass and the size of a progenitor disk are $M_{\rm d}$
 and $R_{\rm d}$, respectively. 
 From now on, all the mass and length are measured in units of
  $M_{\rm d}$ and  $R_{\rm d}$, respectively, unless specified. 
  Velocity and time are 
  measured in units of $v$ = $ (GM_{\rm d}/R_{\rm d})^{1/2}$ and
  $t_{\rm dyn}$ = $(R_{\rm d}^{3}/GM_{\rm d})^{1/2}$, respectively,
  where $G$ is the gravitational constant and assumed to be 1.0
  in the present study. 
  If we adopt $M_{\rm d}$ = 6.0 $\times$ $10^{10}$ $ \rm M_{\odot}$ and
  $R_{\rm d}$ = 17.5 kpc as a fiducial value, then $v$ = 1.21 $\times$
  $10^{2}$ km/s  and  $t_{\rm dyn}$ = 1.41 $\times$ $10^{8}$ yr,
  respectively.
  In the present model, the rotation curve becomes nearly flat
  at  0.35   with the maximum rotational velocity $v_{\rm m}$ = 1.8 in
  our units ($\sim$ 220 km/s).
  The corresponding total mass $M_{\rm t}$ and halo mass $M_{\rm h}$
  are 5.0  and 4.0 in our units, respectively.
  The radial ($R$) and vertical ($Z$) density profile 
  of a  disk are  assumed to be
  proportional to $\exp (-R/R_{0}) $ with scale length $R_{0}$ = 0.2
  and to  ${\rm sech}^2 (Z/Z_{0})$ with scale length $Z_{0}$ = 0.04
  in our units,
  respectively.
  The velocity dispersion
  of halo component 
   at a given point
  is set to be isotropic and given
  according to the  virial theorem.
  In addition to the rotational velocity made by the gravitational
  field of disk and halo component, the initial radial and azimuthal velocity
  dispersion are given to disk component according
  to the epicyclic theory with Toomre's parameter (\cite{bt87}) $Q$ = 1.2.
  The vertical velocity dispersion at given radius 
  are set to be 0.5 times as large as
  the radial velocity dispersion at that point, 
  as is consistent  with 
  the observed trend  of the Milky Way (e.g., Wielen 1977).
 As is described above, the present initial disk model does not
 include any remarkable bulge components, and accordingly corresponds to
 `purely'  late-type spiral without galactic bulge. Although it is
 highly possible that  galactic bulges greatly affect the chemical evolution
 of galaxy mergers, we however investigate
 this issue in our future papers.

\subsection{Gas, star formation, and chemical enrichment}

  The collisional and dissipative nature 
  of the interstellar medium is  modeled by the sticky particle method
  (\cite{sch81}).
  We assume that the fraction of gas mass ($f_{g}$) in
  a disk is set to be 0.2 initially.
  The radial and tangential restitution coefficient for cloud-cloud
  collisions are
  set to be 1.0 and
  0.0, respectively.
  Star formation
   is modeled by converting  the collisional
  gas particles
  into  collisionless new stellar particles according to the algorithm
  of star formation  described below.
  We adopt the Schmidt law (Schmidt 1959)
  with exponent $\gamma$ = 2.0 (1.0  $ < $  $\gamma$
      $ < $ 2.0, \cite{ken89}) as the controlling
    parameter of the rate of star formation.
The positions and velocity of the new stellar particles are set to 
be the same as those of original gas particles.
 Chemical enrichment through star formation during galaxy merging
is assumed to proceed 
both locally and instantaneously in the present study.
The fraction of gas returned to interstellar medium
in each stellar particle and the chemical yield 
are
0.3 and 0.02, respectively.
  The  number of particles 
  for an above  isolated galaxy is 
  5000 for dark halo,   
5000 for stellar disk components, and 5000 for gaseous ones.
Therefore in total  75000 particles are used in a simulation
of multiple merging in the present study.

\subsection{Initial conditions of multiple mergers}

We investigate two different sets of multiple merger models.
For one sets of models,
the initial position of
each progenitor disk (i.e., the center of mass of the disk)
is set to be distributed
randomly within a sphere with the radius of 8.0 in
our units (corresponding to 140 kpc), 
and the initial three-dimensional
velocity dispersion of each disk (that is,
the random motion of each galaxy in the sphere) is set to be
distributed  in such a
way that the ratio of the total kinematical
energy to the total potential energy in the system is  $t_{v}$.
The $t_{v}$ is assumed to be
a free parameter that controls
the degree of dynamical equilibrium
(i.e., $t_{v}$ = 0.5 corresponds to virial equilibrium),
and the results of the models
with $t_{v}$ = 0, 0.1, 0.3 are described in the present study.
For the model with larger $t_{v}$, the time scale for which
all five disks merge to form a single elliptical galaxy 
is longer.
This set of models is hereafter referred to as the 3D model.
For the other set of models,
the initial position of
the center of mass of the disk
is set to be distributed
randomly within a plane ($x$-$y$ plane).
All of the  centers  of mass of the five disks
are located within a circle (on $x$-$y$ plane) 
with the radius of 8.0 in
our units (corresponding to 140 kpc).
The initial orbital planes  of all disks
are assumed to be exactly the same as the $x$-$y$ plane.
The initial two-dimensional velocity dispersion of each disk (that is,
the random motion of each galaxy in the plane) is set to be
distributed  in such a
way that the ratio of the total kinematical
energy to the total potential energy in the system is  $t_{v}$.
Physical meaning of $t_{v}$ in this set of models is exactly
the same as that for the 3D models.
This set of models is  hereafter referred to as the 2D model.
The time when the progenitor disks merge completely and reach  the
dynamical equilibrium is less than 20.0 in our units for most of
models.
For both sets of models, initial internal spin vectors
of five disks are  distributed randomly.
By comparing the results of the 3D models
and those of the 2D ones, we can grasp some essential ingredients
of star formation histories  of multiple galaxy mergers.  
We present numerical results of three models
with $t_{v}$ = 0, 0.1, 0.3 both for the two sets of models. 
The 3D model with  $t_{v}$ = 0 is referred to as the standard model.

   All the calculations related to 
the above dynamical evolution  including the dissipative
dynamics, star formation, and gravitational interaction between collisionless
and collisional component 
 have been carried out on the GRAPE board
   (\cite{sug90})
   at Astronomical Institute of Tohoku University.
   The parameter of gravitational softening is set to be fixed at 0.03  
   in all the simulations. The time integration of the equation of motion
   is performed by using 2-order
   leap-flog method. Energy and angular momentum  are conserved
within 1 percent accuracy in a test collisionless merger simulation.
Most of the  calculations are set to be stopped at T = 20.0 in our units
unless specified.

\placefigure{fig-2}
\placefigure{fig-3}

\section{Result}
\subsection{The standard model}
 We first describe the results of 
the standard model,
which shows typical behaviors in star formation
history of multiple galaxy mergers.
Owing to the initial very small velocity dispersion
of galaxies in this model, all five disks rapidly merge 
to form a single elliptical galaxy within 1.7 Gyr.
The detailed morphological, structural, and kinematical properties
of remnants of multiple galaxy mergers are given in
Barnes (1989), Weil \& Hernquist (1994, 1996).
Therefore we here concentrate on the star formation history
and the chemical evolution of multiple mergers. 
In the followings,  $T$ (in units of Gyr) 
represents the time that has elapsed since the five disks begin
to merge.
For convenience, collisionless
stellar components that are initially
gaseous one and later formed
by star formation are referred to as new star(s) whereas
those that are initially stellar components at $T$ = 0
is referred to as old  star(s) or simply star(s). 

\placefigure{fig-1}
\placefigure{fig-2}
\placefigure{fig-3}
\placefigure{fig-4}
\placefigure{fig-5}
\placefigure{fig-6}

\subsubsection{Star formation history}

Figure 1, 2, 3, and 4 show the time evolution of
dark matter halo, star, gas, and new star, respectively,
for the standard model. Owing to the lack of 
a common dark matter halo, 
the strong dynamical friction 
drives the five individual halos
of disk galaxies to merge very quickly with each other   
to form a single large halo within $\sim$ 1.7 Gyr.
Two equal-mass
disks first interact strongly with each other at $T$ = 0.14 Gyr
(hereafter in units of Gyr)
and consequently forms a long tidal tail at $T$ = 0.28
(See also Figure 5).
Before these two disk completely merge, another two disks
begin to interact with the merging two disks at $T$ = 0.4.
As the merging between these four galaxies proceeds,
the long tidal tail formed in the first tidal encounter
is destroyed and dispersed into the surrounding intergalactic 
regions at $T$ = 0.62. 
Another weak tidal tails are then  developed when the four galaxies
finally merge to form a single elliptical galaxy at $T$ = 0.62.
The surviving  one disk galaxy, which is relatively undisturbed
in the merger events before $T$ $<$ 0.62,
 also finally merges  with the previously merged disks
and consequently forms a single elliptical galaxy till $T$ = 1.69.
A very long and remarkable tidal tail with the projected length 
of $\sim$ 100 kpc is developed during this final phase of
multiple galaxy merging.
In the present model, the morphological
properties of tidal tails formed in multiple merging
are more complicated  and the number of 
distinct tails is larger compared with the case of pair
merging (between only two disks).

As is shown in Figure 6,  strong starbursts with the star formation
rate of 60 $\sim$ 120 $M_{\odot}$ ${\rm yr}^{-1}$ occurs three times
during galaxy merging (0 $<$ $T$ $<$ 1.5).
The interval between each of the  peaks of the 
star formation ($T$ = 0.40, 0.62, and 1.00)
is about 0.2 $\sim$ 0.4 Gyr, which reflects
the time interval between each of strong interaction/merging
occurred in the present model.
The derived star formation rate corresponds to that required for
explaining the very high infrared luminosity of ULIRGs with
dusty starburst (Sanders \& Mirabel 1996). 
This repetitive massive starburst is one of characteristics of
the star formation histories of multiple galaxy merging
and thus is in striking contrast to the star formation
histories of pair merging: Only  one or two times massive starbursts
are demonstrated to occur in pair mergers (Mihos \& Hernquist 1996).
Star formation rate does not drop so dramatically between
starbursts (i.e., 0.4 $<$ $T$ $<$ 0.62 and 0.62 $<$ $T$ $<$ 1.0)
and keeps  rather high ($>$ 10 $M_{\odot}$ ${\rm yr}^{-1}$)
compared with that of the isolated disk model. 
This is essentially because when the magnitude of a starburst 
triggered in a merger between two disks becomes small 
owing to the rapid gas consumption by star formation, 
another strong interaction and merging
begins and then triggers the next starburst(s) in the standard model.
Because of the repetitive starbursts, about 75 \% of initial
gas mass is found to be consumed, as is shown in Figure 6.
The total amount of gas is nearly the same as that derived 
for pair galaxy mergers (e.g., Mihos \& Hernquist 1996).

Although the time evolution of star formation rate in multiple
mergers can be
different from that of pair ones,
the physical processes that trigger massive starburst
in multiple mergers are nearly the same as those
in pair ones.
As two galaxies first encounter one another in the early phase
of multiple merging ($T$ $<$ 0.4),
the two galaxies are tidally distorted to form strong non-axisymmetric
potential (such as barred potential).
This potential acts to torque the interstellar gas of the two disks
and consequently drives  the rapid inward gas transfer that
is indispensable for massive starbursts.
Since 
initial disks are assumed to have no massive bulges
in the present model,
the triggered starburst can be seen well before the completion
of galaxy merging of the two disks. 
As the galaxies approach very closely with each other and suffer from
violent relation in the late phase
of multiple merging, the rapidly changing gravitational potential
greatly enhances the cloud-cloud collision rate.
As a result of this, gaseous energy dissipation becomes very efficient 
and consequently a large amount of gas can be fueled to the central
starburst regions.
These processes have been already given in detail by Mihos \& Hernquist 
(1996). 
We thus suggest that fundamentally important physical processes
of starbursts in multiple galaxy merging can be basically understand
in terms of the already clarified mechanism of starbursts in pair mergers.

\placefigure{fig-7}
\placefigure{fig-8}
\placefigure{fig-9}
\placefigure{fig-10}

\subsubsection{Chemical properties}

Figure 7 shows age and metallicity distributions for new stars
at $T$ = 1.69 when five disks nearly complete the formation
of  a single elliptical
galaxy.
Three peaks in the age  distribution can be clearly seen,
which reflects the fact that massive starburst occurs three times
during multiple galaxy merging.
Compared with the peculiar age distribution,
only one peak can be seen in the metallicity distribution,
which means that typical metallicity
is not so different between young starburst components.
The peak value of the metallicity distribution is rather high  
($\sim$ 0.05 corresponding to 2.5 times $Z_{\odot}$),
basically because chemical enrichment in
interstellar gas during starbursts proceeds so rapidly and
efficiently that new stars formed from the enriched gas have
relatively large metallicity.
Starbursts triggered by multiple merging thus 
play vital roles in the formation of metal-rich young
stellar populations in elliptical galaxies: Dynamical processes
such as those triggering starburst are important
determinant for chemical evolution
of galaxies and thus  for
the nature of stellar populations of galaxies.
As is described above, young stellar components
are very metal-rich and have 
bimodal or multi-modal age distribution.
This result will provide a clue to the origin
of age and metallicity distributions of 
globular clusters observed in 
elliptical galaxies and some merging ones (e.g., Whitmore 1997).

Tidal stripping of gaseous components,
which causes  inhomogeneous chemical mixing, 
can greatly affect
the gaseous  chemical evolution
during multiple galaxy merging. 
Figure 8 shows gaseous metallicity distribution
of the merger remnant of the standard model
for the outer halo region ($R$ $>$ 17.5 kpc corresponding to
the initial disk size, where $R$ is the distance from
the center of mass of the remnant)
and for the whole region.
The peak value in metallicity distribution  
both for the outer halo and for the whole region
is rather small ($\sim$ 0.01).
In particular, a significant fraction of gaseous components ($\sim$ 30 \%) 
is found to
show sub-solar metallicity (less than 0.01) for
the outer halo region.
These results imply that an elliptical galaxy formed
by multiple galaxy merging has
the outer  metal-poor gaseous halo with the mean metallicity of sub-solar.
Figure 9 and 10 describe the two-dimensional distribution
of metallicity at $T$ = 1.69  for stellar components and 
that for gaseous ones, respectively.
Both stellar metallicity and gaseous one are larger in the central
part than in the outer one, which means that the merger remnant
shows negative metallicity gradients both in stellar  
components and in gaseous ones.
What is the most significant in this figure is
that the remnant has the metal-poor gaseous halo 
with the metallicity of  $\sim$ 0.01 (also
relatively metal-poor diffuse stellar halo).
The implications of the derived metal-poor gaseous
halo are given later in the section of discussion (\S 4).

The details of the  formation process of gaseous metallicity gradients 
are given  as 
follows.
In the present chemodynamical model of gas-rich galaxy mergers,
metals produced and ejected in star-formating gaseous 
regions of a multiple merger 
can be mixed only locally into the surrounding ISM (`Inhomogeneous
chemical mixing').
Accordingly, the metals, which are mostly produced in the central region
of the  merger,  can be mixed preferentially into 
ISM  in the central region where further efficient star formation
is expected and  thus can not be mixed so efficiently into
the outer region of the merger.
Consequently, the ISM  of the outer part of the
merger remains less metal-enriched.
Such less metal-enriched ISM in the outer part of the merger 
is   then  effectively stripped away from the system during tidal interaction
of galaxy merging and finally  transferred to the more outer region
where  metals produced in the central part of the  merger
are harder to be mixed into.  
As a natural result of this, the mean metallicity 
of the ISM remaining in the outer part of the remnant
becomes   considerably smaller.
On the other hand, the ISM initially located in the central part of
the  merger 
can be more metal-enriched 
owing to quite efficient star formation there. 
The formation of gaseous metallicity gradients 
accordingly reflects  the fact
that the  metals produced by star formation  
are more efficiently trapped by the ISM in the central part of mergers than by
that  in the outer part.
Thus, the origin of negative metallicity gradients in the gaseous halos of  merger remnants
is due principally to the inhomogeneous chemical mixing in gas-rich mergers.

Importance of tidal stripping in chemical evolution of galaxy mergers
are first investigated by Bekki (1998), though the range of the parameters
for the orbital configurations and the number of merger progenitor disks
is very narrow (Only two models are investigated in this previous paper).
We confirm that the importance of tidal stripping is not dependent on
model parameters such as $t_v$. In particular,
metal-poor gaseous halo surrounding an elliptical galaxy formed
by multiple merging is found to be seen for all models in the present
study. We discuss later the formation of metal-poor halo
in the context of a fossil record of the past merging events
for field elliptical galaxies.

\placefigure{fig-11}
\placefigure{fig-12}

\subsection{Parameter dependences}

Figure 11 describes the dependence of the
time evolution of star formation history 
and gas mass fraction on 
the parameter $t_v$ for the 3D model. 
We can sen the following four  clear dependences
in this figure.
Firstly, irrespectively
of $t_v$, strong starbursts with the star formation rate larger than
10 $M_{\odot}$ ${\rm yr}^{-1}$ occur in a repetitive way. 
This is basically because the five disks in a model
do not merge with each other at the same time and consequently
experience two or three times strong dynamical interaction and merging.
Secondly,  the magnitude of the maximum starburst is 
larger for the model with smaller $t_v$.
The reason for this dependence
is described as follows.
Owing to the initial smaller  velocity dispersion,
five disks can interact dynamically with each other
in the earlier merger phase when the disks have a larger amount
of interstellar gas.
As a result of this, a larger amount
of gas can be transferred to the central region during multiple merging.
Furthermore  a smaller amount of the gas
is tidally stripped away from the merger in the model
with smaller $t_v$. 
Thus a larger amount of gas accumulated within the central
region of the merger can be consumed by the central massive starbursts. 
Thirdly the time interval between the first starburst and the last
one is longer  for the model with larger $t_v$,
which  reflects the fact that 
final merging that produces a single elliptical galaxy
occurs later in the model with larger $t_v$. 
Fourthly, the final gas mass fraction is not so greatly different
between the three models, though the strength and the epoch
of massive starburst are different between these models:
About 70 \% $\sim$ 75 \% of initial gas can be consumed by star formation
in multiple mergers.
This result is very similar to that of pair mergers revealed
by Mihos \& Hernquist (1996).

Figure 12 describes the dependence of the time evolution
of star formation history and gas mass fraction on $t_v$
for the 2D model.
There are some differences in the $t_v$ dependences between
the 2D model and the 3D one.
Firstly, the triggered starbursts are a factor
of 2 $\sim$ 3 stronger in the 2D model
than in the 3D model.
The first reason  for this is that
owing to the stronger dynamical friction,
the five disks can merge at roughly the same time
so that 
the larger amount of gaseous components  can collide with each other,
dissipate their kinetic energy,  be transferred to
the central region,
and be consumed very efficiently by star formation. 
The second reason is that initial orbits of the five disks in the 2D model
is confined in $x$-$y$ plane, a smaller amount of gas indispensable
for massive starbursts 
can be stripped away from the merger.
Secondly, the repetitive starburst is not so clearly seen in 
the 2D model: Only the 2D model with $t_v$ = 0.3 shows two times
starbursts.
This result clearly reflects
the fact that 
most of the five disks merge with each other at roughly
the same time so that most of the gas can be consumed
by the first (and the last) merging 
in the 2D model.
Thirdly, the magnitude of the maximum starburst
is not necessarily larger in the model with smaller $t_v$:
Although the starburst strength is larger in the models with
$t_v$ = 0 and 0.1
than in that with $t_v$ = 0.3, the strength
is larger in $t_v$ = 0.1 than in $t_v$ = 0.
This result suggests that
in addition to 
$t_v$, there are some important parameters that control the
strength of starbursts in the 2D model.
We suggest that in the 2D model, orbital configurations
of galaxy merging (e.g.,  whether a merger is a
prograde-prograde merger  or a retrograde-retrograde one)
is also a determinant for the strength of
the maximum starburst triggered by multiple galaxy merging.

Figure 13, 14, 15, and 16 summarize the morphological properties
of multiple mergers at the epoch of the starburst for
the 3D models with $t_v$ = 0.1 and 0.3 and the 2D model
with $t_v$ = 0.
The maximum star formation rate is found to range from
$\sim$ 60 $M_{\odot}$ ${\rm yr}^{-1}$ to $\sim$ 300  
$M_{\odot}$ ${\rm yr}^{-1}$ in the present study.
Accordingly, the derived morphological properties
are useful and helpful for understanding the nature
of infrared luminous galaxies and ULIRGs in which
star formation rate is required to be the order of 10 $\sim$ $10^2$ 
$M_{\odot}$ ${\rm yr}^{-1}$ for the large infrared luminosity
(if the source of the reemission is due mostly to dusty starburst),
though the parameter range investigated in the present study
is relatively narrow. 
As is shown in Figure 13,   the global morphology  
of the multiple merger is very complicated particularly
for the early epoch of merging (i.e., the first starburst epoch,
$T$ = 0.59):
Three long tidal tails, multiple nuclei composed mainly
of young stars, and a disk (seen from edge-on view) that
is not largely disturbed can be seen at the epoch of maximum starburst.
Probably a trend that  both few strong tidal tails  
and an undisturbed disk can be seen 
in a merger
is specific for multiple mergers (i.e., not seen in pair ones). 
This trend is true for most of the 3D models  
in the earlier  starburst epoch.
As is shown in Figure 14, the global morphology 
at the last starburst epoch ($T$ = 1.74) in the merger model with
$t_v$ = 0.3 does not appear to be so disturbed as that in Figure 13 and 16. 
This is essentially because in the last merging,
most of the remarkable tidal tails that are formed in the early
encounter and merging can be greatly dispersed to form outer
diffuse stellar and gaseous halo.  
What is remarkable in this model is that 
although the global
morphology of the merger looks like a single elliptical galaxy,
the central part of the merger (the central 4.38 kpc) shows  distinct
two cores composed mostly of new stars (See Figure 15).
The 2D model with $t_v$ = 0 is also found to show very
complicated tidal tails at the epoch of the maximum starburst (See Figure 16).
What is interesting in this model is that the morphology
of young stellar components looks like a tadpole 
in edge-on view and thus is similar
to the morphology of Mrk 273, which is an ULIRG.
Thus morphological properties of multiple mergers
at the epoch of their massive starbursts
 are found to
depend on model parameters and thus to be  
very diverse.
Recent observational studies based on
$HST$
and large grand-based telescopes  have revealed
the great difference in structural and morphological
properties of ULIRGs (Surace et al. 1998;
Dinshaw et al. 1999; Surace, Sanders, \& Evans 1999;
Scoville et al. 1999)
For example,
observational studies based on the  NICMOS imaging of ULIRGs have
found that the morphologies are rather diverse ranging
from exponential disks  with long tidal tails to
only one core with the $R^{1/4}$ law density profiles
(e.g., Scoville et al. 1999).
The present
 results  imply that the observed diversity in morphological properties
of ULIRGs can be partly understood in terms of the difference of  physical
parameters between multiple mergers.

\placefigure{fig-13}
\placefigure{fig-14}
\placefigure{fig-15}
\placefigure{fig-16}

\section{Discussion}

\subsection{Star formation history in a compact group of galaxies}
The present numerical model has  demonstrated that  
multiple tidal interaction and merging can trigger
massive starbursts and the maximum star formation
rate can reach $\sim$ 100 $M_{\odot}$ ${\rm yr}^{-1}$
that is required for explaining the very large infrared luminosity
of typical ULIRGs.
Since multiple galaxy merging is suggested to occur naturally
in compact group of galaxies (Barnes 1989; Hickson 1997), 
we here discuss the derived results
in the context of the evolution of galaxies in compact group
of galaxies. 

\subsubsection{Strength of starburst}
First problem is on whether a massive starburst 
($\sim$ 100 $M_{\odot}$ ${\rm yr}^{-1}$) can be triggered
in some member galaxies
during  the dynamical evolution of a compact group of galaxies.
We here stress that the derived star formation rate is
rather overestimated owing to the following two assumptions adopted
in the present study.
Firstly, for simplicity,  we assumed that all five disks have the
same gas mass fraction ($f_{g}$) of 0.2.  It is suggested that if
density-dependent star formation law is applied (i.e., the Schmidt law),
initial gas mass fraction of merger progenitor disk is a
critically important factor for the strong starburst (Bekki 2000).
Therefore, if a multiple merger has a few gas-poor disks ($f_{g}$ much
less than
0.2),
the strength of the triggered starburst is considerably smaller
compared with multiple mergers with all disks being gas-rich
(i.e., the mergers investigated in the present study).
Secondly, we assumed that the initial disk mass is the same
for all merger progenitor galaxies: The massive starburst
is basically triggered by major galaxy merging between  equal-mass disks
in the present study.
Although minor and unequal-mass galaxy merging can trigger
starburst (Mihos \& Hernquist 1994), the magnitude of the starburst
is rather small compared with major galaxy merging (Bekki 1998).
Therefore, if there is a significant difference in mass between
merger progenitor disks, the triggered starburst becomes very
small.

Although it is very difficult to discuss what the most probable
initial physical  conditions of gas mass fraction and mass ratio of galaxies
are for multiple galaxy mergers, 
we can address this issue by following recent observational results
on galaxies in compact groups.
Williams \& van Gorkom (1988) and Williams et al. (1991)
showed that the cool gas is not confined to the member galaxies
of compact groups and accordingly suggested that
many compact group evolved to the point that cool gas contained
within individual galaxies can be so efficiently distributed
throughout the group (owing to tidal stripping
of the gas) that the gas fraction of individual
galaxies becomes very small. 
The fraction of cool interstellar gas estimated
from CO-line observation is found to
be similar to that of isolated spiral galaxies (Boselli et al. 1996),
though this result may be biased by the relatively small size of
compact group of galaxies (Hickson 1997).
The fraction of gas-rich late-type disks is demonstrated
to be significantly less ($<$ 0.5) in compact group than in the field
(e.g., Hickson 1982; Sulentic 1987; Hickson et al. 1988).
These observational results seem to imply that
the mean gas mass fraction of galaxies in compact groups
is rather small (i.e., the strength of starburst derived in
the present study is overestimated).
We here stress that the above observational results and
their implications are true only
for low redshift compact group of galaxies and therefore that
the gas mass fraction of merging compact group of galaxies
can be very high if the merging can occur at high redshift.

Distribution of relative galaxy luminosity (i.e.,
the luminosity difference between the brightest galaxy
and other member galaxy in a compact group)
can provide a clue to the question on the above second assumption
(Hickson 1982). 
Hickson (1982) investigated  the difference in  galaxy luminosity
(indicated in Figure 3 of Hickson 1982) and found that 
the difference is larger for a compact  
group with (first-ranked) elliptical galaxies than for that without.
Figure 3 in  Hickson (1982) clearly showed that
there are a significant fraction
of compact groups with the difference in magnitude (and possibly, mass)
between member galaxies larger than 2.0 mag, 
though  Hickson (1982) did not clearly mention that. 
These results imply that multiple $major$ merging is less likely
in the evolution of compact groups and thus that the strength of 
starburst reported by the present study can be overestimated. 
Thus these observational results on the above our two assumptions
seem to imply that compact group of galaxies are less likely
to trigger massive starburst with the star formation rate of
$\sim$ 100 $M_{\odot}$ ${\rm yr}^{-1}$.
However,
considering the smaller number of the detailed
observational studies on physical properties
of individual galaxies in compact groups,  it is fair
to say that only a compact group in which
the difference in mass between the member galaxies
is very small and most of member galaxies are gas-rich
can trigger massive starburst with the star formation 
rate of $\sim$ 100 $M_{\odot}$ ${\rm yr}^{-1}$. 

\subsubsection{An evolutionary link between ULIRGs and
multiple mergers}

Second question is on whether a compact group of galaxy
can evolve into an ULIRG.
Xia \& Deng (1997) first suggested that
multiple galaxy merging can be closely associated with
the formation of some ULIRGs, based on the observational 
results of spatial distribution of soft X-ray emission in Mrk 273
and the existence of more than 10 dwarf galaxies within 100 kpc
of Mrk 273.
Nishiura et al. (1997) discussed whether HCGs can
be regarded as precursors of ULIRGs by comparing
the $K$ band luminosity function of the HCGs
with that of ULIRGs.
Taniguchi \& Shioya (1998) suggested that 
Arp 220, which is a typical  ULIRG,
is formed by multiple galaxy merging, based on
the total number of OH maser sources and its spatial distribution
in the central region of Arp 220.
Recent morphological studies by the Hubble Space
Telescope ($HST$) have found that 22 among 99 ULIRGs 
in the redshift range 0.05 $<$ $z$ $<$ 0.20 have
possible multiple nuclei (Borne et al. 2000).
Based on these new results,
Borne et al. (2000) suggested  that
multiple merging not only plays decisive roles
in the formation of ULIRGs but also transforms
a compact group of galaxies into a single elliptical galaxy. 
The idea that ULIRGs are formed by multiple merging
is not fundamentally new in the sense that  Sanders et al. (1998)
have already suggested how dissipative galaxy merging
is important for the formation of ULIRGs
and Mihos \& Hernquist (1996) have already 
revealed that efficient inward transfer of gas can trigger
massive starburst in galaxy mergers:
Irrespectively of whether a merger is a pair or a multiple,
essentially important physical processes associated 
with the formation of starburst is probably
the same between the two apparently
different mergers.
The  difference between the model by Sanders et al. (1988) 
and the multiple merger one 
by the above authors is only the number of galaxies
in merging.
Accordingly  it seems to be nearly meaningless to 
discriminate clearly the two models (pair merger model
and multiple merger one).
However we here consider
that it is  important to investigate observationally what fraction
of ULIRGs are formed by multiple galaxy merging.
This is 
because the fraction enables us to estimate (though indirectly)
the fraction of compact groups that are  now evolving
from ULIRGs into field elliptical galaxies.

The observed infrared luminosity of ULIRGs is
successfully reproduced by pair mergers (Bekki et al. 1999),
and furthermore the diversity in morphological properties
between ULIRGs is suggested to reflect the difference
in orbital configurations and internal structure between
pair mergers (Bekki 2000).
The present study on multiple galaxy merging,
on the other hand, demonstrated that   
a multiple merger shows such a high star formation rate  that
it can be identified as an ULIRG, and furthermore that
the morphology  of the merger at the epoch of massive starburst
(i.e., the epoch when the merger can be observed as an ULIRG)
is similar to those of some ULIRGs.
Numerical results on pair mergers (Bekki et al. 1999; Bekki 2000)
and those for multiple
ones (i.e., the resent study)
therefore imply that morphological and photometric
properties alone cannot clearly indicate whether an ULIRG
is a pair merger or a multiple one.
As is suggested by Borne et al. (2000),
one of key observational tests  concerning the above
question is to investigate  whether  
each of apparent multiple nuclei observed in
some  ULIRGs (e.g., those shown in Figure 1 of Borne et al. 2000)  
is really a galactic nucleus or a super-starburst.
Recent high resolution imaging of some ULIRGs has
revealed that most of ULIRGs show compact blue knots of
star formation with the number of such knots ranging from
4 to 31 per object (Surace et al. 1998).
Surace et al. (1998) furthermore 
revealed that the range of the knot masses is $10^{5}$-$10^{9}$ 
and the  upper age limit for the knots in individual
galaxies is $\sim$ 3 $\times$ $10^{8}$ yr 
by using spectral synthesis modeling.
The existence of these  star-forming blue knots have been
confirmed by several other observational studies 
based on high resolution optical/near-infrared imaging
of ULIRGs (Dinshaw et al. 1999; Surace et al. 1999; Scoville et al. 1999). 
Furthermore Bekki (2000) demonstrated that
most of morphological properties of ULIRGs
can be reproduced by pair mergers at the epoch of the maximum starburst
(i.e. the epoch when mergers can be observed as ULIRGs).
Accordingly it is highly possible that ULIRGs with several apparent nuclei 
appear to be pair mergers with super-starburst knots,
though theoretical studies on major mergers
have not yet clarified the mechanism for the formation
of such massive blue knots observed in ULIRGs. 
Here it should be noted
that Barnes \& Hernquist 
(1992b) have already demonstrated that massive dwarf-like
objects can be formed in tidal tails of a pair merger. Accordingly
it is not unreasonable to consider that these newly
formed objects can be identified
as blue knots in ULIRGs.
These observational  and theoretical results
imply that ULIRGs with apparent multiple nuclei
are not  formed by multiple mergers:
A pair galaxy merger  both triggers massive starburst
and forms several very bright massive star clusters
that look like galactic nuclei in optical bands. 
However, high-resolution multi-band studies 
(in particular, in  near-infrared band) on morphological properties 
of ULIRGs have not been so accumulated yet which
can clearly discriminate real galactic nuclei with
very bright star clusters. 
Accordingly it is fair for us to say that it is still
highly uncertain what fraction of ULIRGs are formed
by multiple merging.

\subsection{Fossil records of the past merging events in field
elliptical galaxies} 

We have demonstrated that owing to the strong tidal stripping
of metal-poor interstellar gas initially within the outer part
of disks, an elliptical galaxy formed by multiple galaxy merging
can have metal-poor gaseous halo and show the negative metallicity
gradient.
There may or may not be recent observational evidences
(e.g., those from   
$Advanced$ $Satellite$ $for$ $Cosmology$ $and$ $Astrophysics$,
$ASCA$)
that support the above results. 
Fe abundance of
hot gaseous X-ray halo
has been suggested to be
appreciably  smaller than that of the stellar component in
the host field elliptical galaxy (Awaki et al. 1994;
Matsushita et al. 1994; Matsumoto et al. 1997;
but see Matsushita et al. 1997).
Furthermore, the hot $X$-ray gaseous halo in elliptical galaxies
has been suggested  to show strong negative metallicity gradients,
which implies that some physical mechanisms such
as cooling flow, gaseous dissipation, galaxy merging,
and dilution from external metal-poor gas play
a vital role in the formation of the gaseous metallicity gradients
(Loewenstein et al. 1994; Mushotzky et al. 1994;
Matsushita et al. 1997). 
Since the  above observationally suggested  metal-poor halo
and metallicity gradients can be reproduced by 
the present numerical results, 
we here propose that the metal-poor gaseous halo is
a fossil record of the past multiple merging events
for field elliptical galaxies.
Multiple galaxy merging can not only transform
a compact group of galaxies into an field elliptical galaxy
but also form metal-poor gaseous halo.
Therefore some of field elliptical galaxies
can be  observed to show metal-poor gaseous halo.

One of observational tests to assess the validity of the
above proposal is to investigate in detail the two-dimensional
distribution of metal-poor gaseous halo in field elliptical galaxies.
Metal-poor gas tidally stripped from merger progenitor disks
is initially distributed in a very inhomogeneous way owing to the
incompletion of dynamical relaxation of multiple galaxy merging.
For a  merger (or a merger remnant) with stronger tidal
disturbance, the meta-poor gaseous halo can be distributed in 
a more inhomogeneous way around the merger. 
Therefore it is highly possible that a field elliptical galaxy
just formed by merging shows very inhomogeneous distribution
of metal-poor gaseous halo. 
Furthermore the degree of inhomogeneity in the distribution
of metal-poor gaseous halo for merger remnants 
can depend strongly on 
how the remnants  are dynamically relaxed.
Accordingly we suggest that if future observations
discover very inhomogeneous distribution of 
metal-poor halo in an field elliptical galaxy, 
the observed halo can be interpreted
as a fossil record of the past multiple merging 
that formed  the elliptical.
We furthermore suggest that if there is a 
strong physical correlation between the degree of inhomogeneity
in the distribution of metal-poor gaseous halo
and that of dynamical relaxation for field elliptical galaxies,
the relation strengthens the scenario that $some$ field
elliptical galaxies can be formed by multiple merging
and evolved from compact groups of galaxies. 
Since the origin of field elliptical galaxies
is still highly uncertain (Bender 1996),
future observational studies on the detailed distribution
of the metal-poor gaseous halo will provide
new and valuable information on the formation and the evolution
of field elliptical galaxies.  

\section{Conclusion}

We have numerically investigated 
the star formation history  and the chemical evolution  
of multiple mergers between five identical gas-rich disk
galaxies in an explicitly self-consistent manner.
The main results of the present study are summarized as follows.

(1) We found that repetitive massive starburst
with the star formation rate ranging from  $\sim$ 10 
to $\sim$ $10^{2}$ $M_{\odot}$ ${\rm yr}^{-1}$   
can occur with the time interval between the starbursts
depending on
the model parameters.
The time interval between the epochs of 
repetitive starburst is longer for the merger with the larger virial ratio
(i.e., the ratio of total kinematical
energy to total potential energy).

(2) The magnitude of the starburst is found to depend on initial
virial ratio 
such that the maximum star formation rate
is longer for the merger with the larger virial ratio.

(3) Initial orbital configurations are also found to
be important for the strength of starburst triggered in
multiple merging. The 2D model is likely to show stronger starburst  
than the 3D model during multiple merging.

(4) Morphological properties of multiple mergers at the epoch
of strong starburst are rather diverse: Some mergers show multiple starburst
nuclei and very long tidal tails.

(5) We also found that chemical evolution  during multiple merging
proceeds so inhomogeneously that metal-poor
gaseous halo can form in the outer part of the merger remnants.
We accordingly suggest that metal-poor gaseous halo
in  a field elliptical galaxy is a fossil record of the past
multiple merging events for the galaxy.

Multiple merging is suggested to occur in the dynamical evolution
of compact groups of galaxies (e.g., Mamon 1987; Barnes 1989;
Hickson 1996 for a recent review).
These numerical results thus provide some important implications
on the evolutionary link between compact group of galaxies,
ULIRGs, and field elliptical galaxies.

\acknowledgments
We are  grateful to the anonymous referee for valuable comments,
which contribute to improve the present paper.

\clearpage


\figcaption{
Time evolution of mass distribution projected
onto $x$-$y$ plane for dark halo components 
at each time  $T$ for the standard model.
The time (in units of Gyr)  indicated in the upper
left-hand corner represents the time  that has elapsed
since the five disks begin to merge. 
Here the  scale is given in our units (17.5 kpc) and 
each of the  12  frames measures 192.5 kpc (11 length units) 
on a side.    
\label{fig-1}}

\figcaption{The same as Figure 1 but for old stars that
are initially collisionless stellar components. 
Star formation  rate reaches  the peak value at $T$ = 0.40, 0.62,
and 1.00 Gyr in this model. The morphology at each of these epochs
is given in the corresponding frame. 
\label{fig-2}}

\figcaption{The same as Figure 1 but for gas. 
\label{fig-3}}

\figcaption{The same as Figure 1 but for new stars. 
\label{fig-4}}

\figcaption{The same as Figure 2  but for $x$-$z$ plane.
\label{fig-5}}

\figcaption{
Time evolution of star formation rate  
in units of $M_{\odot}$ ${\rm yr}^{-1}$ for  the standard model. 
For comparison, the result of the isolated disk model
is also given. Note that repetitive massive starburst occurs
in the multiple merger model. Note also that 
initial gas is more rapidly consumed by the triggered starbursts
in the merger model than
in the disk one. 
\label{fig-6}}

\figcaption{
Age distribution (upper) and metallicity one (lower) for new stars
in the standard model at $T$ = 1.69 Gyr.
\label{fig-7}}

\figcaption{
Gaseous metallicity distribution of the standard model
at $T$ = 1.69 Gyr for the whole region
(red) and the outer region (blue) 
with $R$ larger than 17.5 kpc (1 length units),
where $R$ is the distance from the center of mass of the merger remnant. 
Here the metallicity distribution is represented by the number
fraction in each metallicity bin. 
Note that a larger fraction of gas shows  sub-solar metallicity (less than
0.02) in the outer part of the merger remnant.
\label{fig-8}}

\figcaption{
Two-dimensional distribution of stellar metallicity (including
both old stars and new ones) 
projected onto $x$-$y$ plane at  $T$ = 1.69  Gyr in the standard model.
The  frame measures  100 kpc on a side and includes 400  bins (20 $\times$ 20).
For the bin within which no stellar particles are found to be located,
any color contours are not given  for clarity.
As is shown in the color legend of this figure,
the metallicity  ranges from 0.01
in the outer part 
to 0.05 in the central one.
Note that this merger remnant shows a negative metallicity
gradient in stellar components.
\label{fig-9}}

\figcaption{The same as Figure 9 but for gaseous components.
\label{fig-10}}

\figcaption{The same as Figure 6  but for 
the 3D models with $t_v$ = 0 (blue), 0.1 (red), and 0.3 (green).
This figure describes the $t_v$ dependences of
the time evolution of star formation history and gas mass fraction.
\label{fig-11}}

\figcaption{The same for Figure 11 but for the 2D models.
\label{fig-12}}

\figcaption{
Mass distribution at the epoch of massive starburst ($T$ = 0.59 Gyr) 
projected onto $x$-$y$ plane (upper three frames)
and $x$-$z$ one (lower three) in
the 3D model with $t_v$ = 0.1 for (old) star (left), gas (middle),
and new star (right).
Each of the six  frames measures   192.5 kpc on a side 
\label{fig-13}}

\figcaption{
The same as Figure 13 but for the 3D model with $t_v$ = 0.3
at the final starburst ($T$ = 1.74 Gyr).
\label{fig-14}}

\figcaption{The same as Figure 14 but for the central region
of the merger. 
Each of the six  frames measures  87.5 kpc on a side.
\label{fig-15}}

\figcaption{
The same as Figure 13 but for the 2D model with $t_v$ = 0
($T$ = 0.46 Gyr).
\label{fig-17}}

\end{document}